\documentclass[times,sort&compress,3p]{elsarticle}
\makeatletter
\def\ps@pprintTitle{%
  \let\@oddhead\@empty
  \let\@evenhead\@empty
  \def\@oddfoot{\reset@font\hfil\thepage\hfil}
  \let\@evenfoot\@oddfoot
}
\makeatother

\journal{Journal of Multivariate Analysis}
\usepackage[labelfont=bf]{caption}

\usepackage{amsmath,amsfonts,amssymb,amsthm,booktabs,color,epsfig,graphicx,hyperref,url}

\theoremstyle{plain}

\theoremstyle{definition}

\usepackage{float}
\usepackage{lscape}

\begin{document}

\begin{frontmatter}

\title{Ethnic Disparities of Female Infiltrating Duct and Lobular Breast Cancer Survival by Cancer Stage: Findings From SEER 2006-2010}

\author{Ishmael Nii Amartei Amartey}

\address{Department of Statistics and Actuarial Science\\ Norhtern Illinois University, Dekalb, IL, 60115 \\ iamartey1@niu.edu \\
 May 08, 2023}


\begin{abstract}

Breast Cancer is a major disease affecting women's health in the United States with incidence and prevalence dominant among younger women and the Black race. We analyzed the association between breast cancer characteristics with age and race and how survival months and age differ in racial groups. Using the Surveillance, Epidemiology, and End Results (SEER) datasets we performed a logistic regression to examine significant predictors that affect survival month. There were 3414 whites, 291 Blacks, and 320 Others (American Indian/AK Native, Asian/Pacific Islander) in the sample. We found significant associations between racial groups and ages with significant differences in age between Blacks and Whites, and Whites and Others.
Patients with a breast cancer tumor in grades 1 and 2 have higher survival months (by 1.49 \% and 0.49\% respectively.
 
\end{abstract}


\end{frontmatter}

\section{Introduction\label{sec:1}}
Breast cancer is one of the commonest cancers in the world accounting for about 12.5\% of all new cancer cases worldwide \cite{breastcancerfactsandstatistics2023} . Statistics from the American Cancer Society \cite{americancancersociety2023} report that breast cancer contributes to about 30\% of all female cancer cases recorded in the US accounting for a ratio of 1:3 of every cancer case.
Despite the significant improvement in research and diagnosis, the disease incidence is still high amongst racial and age groups \cite{foy2018disparities,wojcik1998breast} with Blacks having the highest incidence compared to whites and other racial groups \cite{iqbal2015differences, shoemaker2018differences}. Other studies found that compared to non-Hispanic White,  non-Hispanic Black, Asian American/Pacific Islander, Native American, and Hispanic women have a higher percentage of invasive breast cancers at younger ages and at more advanced stages, and a higher percentage of breast cancer deaths at younger ages \cite{hendrick2021agedistributions}.
In this study, we sought to examine the differences in survival months, regional nodes examined (RNE), positive regional nodes (RNP), age, and tumor size across racial groups of Blacks, Whites, and Others (American Indian/AK Native, and Asian/Pacific Islander). We also seek to establish the associations between breast cancer characteristics across race and age groups and to determine significant predictors of survivability through logistic regression models.

\section{Methods\label{sec:2}}
\subsection{Data Source}

The Surveillance, Epidemiology, and End Results (SEER) provide useful information on cancer statistics in the United States for researchers. In this study, we used a dataset of 4024 female patients diagnosed with infiltrating duct and lobular breast cancer between 2006-2010 \cite{teng2019seer}. The data frame contains information at the time of diagnosis on age, race, marital status, T-stage, N-stage, 6th stage, differentiate, grade, A Grade, tumor size, estrogen, and progesterone status (negative{/}positive), regional node examined, regional nodes (positive{/} examined), survival months, and status (dead{/}alive). The categories in racial groups are Black, White, and Others (American Indian{/}AK Native, Asian{/}Pacific Islander), marital status as single, married, separate, widowed, divorced, T stage into 4 categories ranging from T1 to T4, N stage ranging from N1 to N3, 6th stage into 5 categories of IIA, IIB, IIIA, IIIB, and IIIC. Differentiation was categorized into 4 main parts: moderate, poor, undifferentiated, and well differentiated, and A stage as regional or distant. We categorized age as 30 to 39 years, 40 to 49 years, 50 to 59 years, and $\geq$ 60 years and categorized survival months into ranges of $\leq$ 20 months, 21-49 months, 50-69 months, 70-89 months, and $\geq$ 90 months.

\subsection{Statistical Analysis}
Continuous variables were summarized using measures of central tendency and dispersion, while categorical variables were summarized using frequency measures. Difference between two means was tested using independent sample t-test and between more than two means using analysis of variance (ANOVA) and Bonferroni Post Hoc test, while difference in proportions was tested using chi square test. Unadjusted and adjusted analysis for the survival prediction were performed using logistic regression. Retrospective survival rate (stage -specific diagnosis to death time) was calculated and compared by ethnicity. Significance was set to 0.05. All the analyses were run using SPSS (V.27).

\section{Results\label{sec:3}}
\subsection{Distribution of the sampled data}
\begin{figure}[h]
    \centering
    \includegraphics {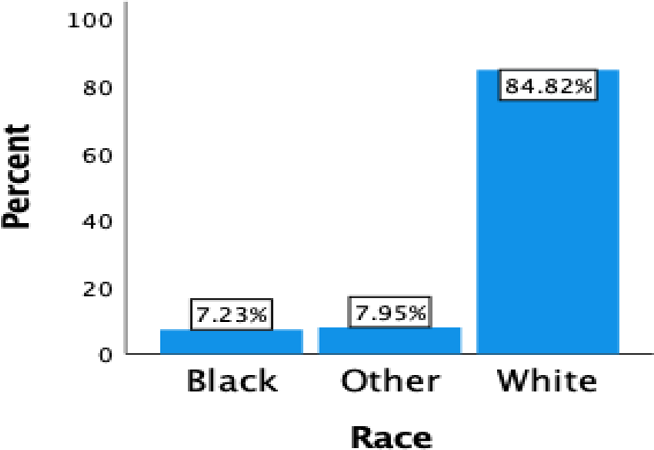}
    \caption{Race distribution of the 4024 sampled women}
    \label{fig1}
\end{figure}

The overall mean and confidence interval (CI) for age, regional node examined (RNE), regional node positive (RNP), tumor size (in millimeters), and survival months as shown in table \ref{table1} were 53.97 years (95\% CI, 53.70 to 54.27), 14.36 (95\% CI, 14.11 to 14.60), 4.42 (95\% CI, 4.00 to 4.32 ), 30.47mm (95\% CI, 29.81 to 31.16), and 71.30 months (95\% CI, 70.58 to 71.95) respectively. The race distribution was 7.23\% blacks, 84.82\%, and 7.95\% others (American Indian/AK Native, Asian/Pacific Islander) as shown in fig \ref{fig1}.

\begin{figure}[h]
    \centering
    \includegraphics[width=15cm]{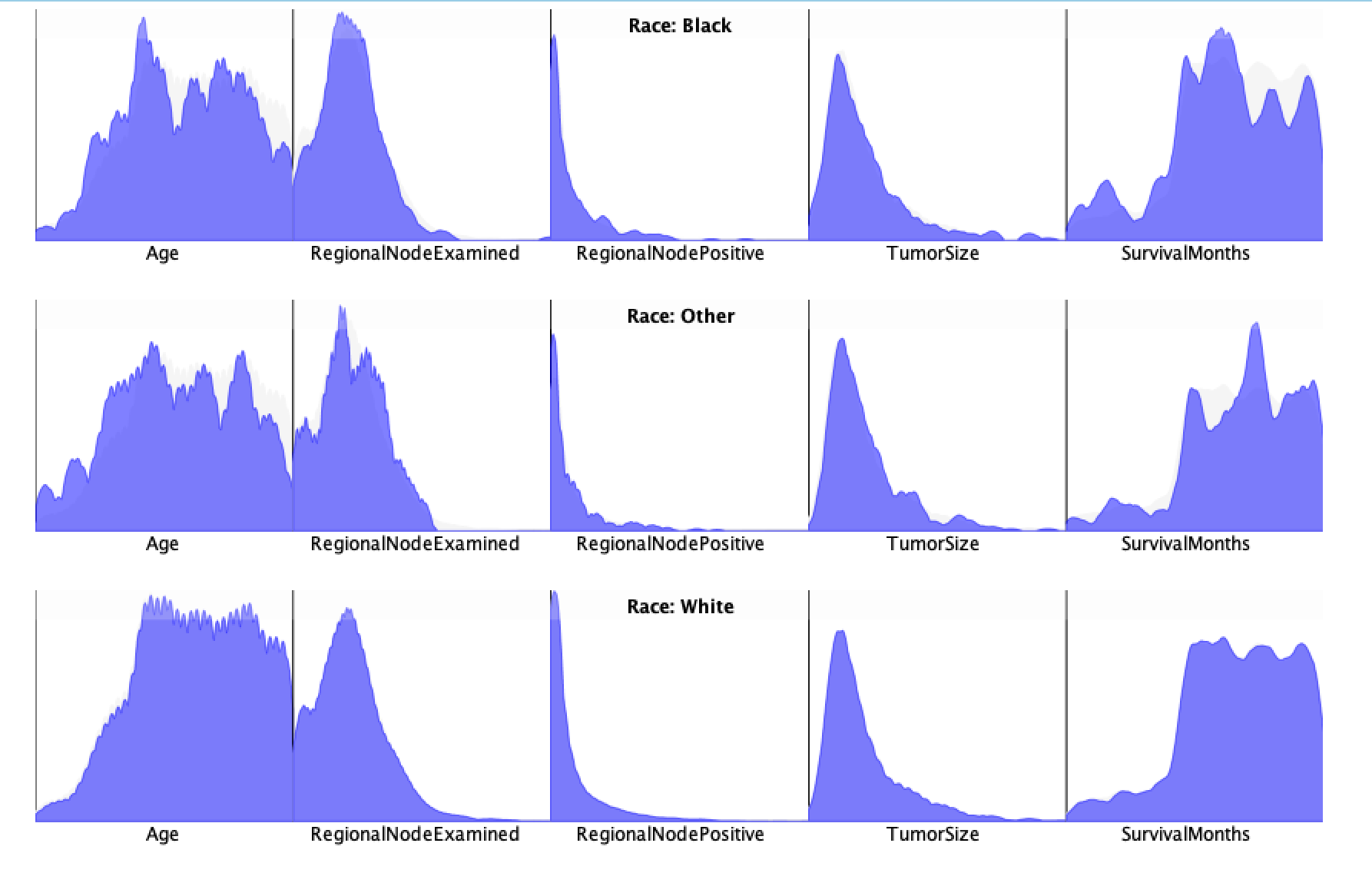}
    \caption{Distribution plot of age, RNE, RNP, Tumor size, and survival months by race}
    \label{fig2}
\end{figure}

\begin{table}[H]
    \centering
    \caption{Distribution of Age, Regional Node Examined (RNE), Regional Node Positive (RNP), Tumor Size, and Survival Months by Race}
    \label{table1}
    \begin{tabular}{lllllll} \hline \hline
Race  &   & Age  & RNE  & RNP & Tumor Size  & Survival Months    \\ \hline
Black & Mean   & 52.58   & 14.31  & 4.42   & 30.52 & 66.61  \\ & n   & 291   & 291   & 291 & 291  & 291        \\ & \begin{tabular}[c]{@{}l@{}}SD\\    \\ 95\% CI\end{tabular} & \begin{tabular}[c]{@{}l@{}}9.050\\    \\ {[}51.62-53.58{]}\end{tabular} & \begin{tabular}[c]{@{}l@{}}7.488\\    \\ {[}13.39-15.12{]}\end{tabular} & \begin{tabular}[c]{@{}l@{}}5.086\\    \\ {[}3.86-5.03{]}\end{tabular} & \begin{tabular}[c]{@{}l@{}}22.167\\    \\ {[}27.98-33.04{]}\end{tabular} & \begin{tabular}[c]{@{}l@{}}24.806\\    \\ {[}63.70-69.55{]}\end{tabular} \\ \hline
Other & Mean                                                       & 51.43                                                                   & 14.71                                                                   & 4.21                                                                  & 31.00                                                                    & 73.22                                                                    \\
      & n                                                          & 320                                                                     & 320                                                                     & 320                                                                   & 320                                                                      & 320                                                                      \\
      & \begin{tabular}[c]{@{}l@{}}SD\\    \\ 95\% CI\end{tabular} & \begin{tabular}[c]{@{}l@{}}9.541\\    \\ {[}50.47-52.51{]}\end{tabular} & \begin{tabular}[c]{@{}l@{}}7.510\\    \\ {[}13.87-15.54{]}\end{tabular} & \begin{tabular}[c]{@{}l@{}}4.897\\    \\ {[}3.68-4.78{]}\end{tabular} & \begin{tabular}[c]{@{}l@{}}19.892\\    \\ {[}29.05-33.39{]}\end{tabular} & \begin{tabular}[c]{@{}l@{}}23.075\\    \\ {[}70.67-75.58{]}\end{tabular} \\ \hline
White & Mean                                                       & 54.33                                                                   & 14.33                                                                   & 4.13                                                                  & 30.42                                                                    & 71.52                                                                    \\
      & n                                                          & 3413                                                                    & 3413                                                                    & 3413                                                                  & 3413                                                                     & 3413                                                                     \\
      & \begin{tabular}[c]{@{}l@{}}SD\\    \\ 95\% CI\end{tabular} & \begin{tabular}[c]{@{}l@{}}8.852\\    \\ {[}54.03-54.64{]}\end{tabular} & \begin{tabular}[c]{@{}l@{}}8.204\\    \\ {[}14.06-14.59{]}\end{tabular} & \begin{tabular}[c]{@{}l@{}}5.131\\    \\ {[}3.96-4.31{]}\end{tabular} & \begin{tabular}[c]{@{}l@{}}21.145\\    \\ {[}29.72-31.14{]}\end{tabular} & \begin{tabular}[c]{@{}l@{}}22.696\\    \\ {[}70.74-72.27{]}\end{tabular} \\ \hline
Total & Mean                                                       & 53.97                                                                   & 14.36                                                                   & 4.16                                                                  & 30.47                                                                    & 71.30                                                                    \\
      & n                                                          & 4024                                                                    & 4024                                                                    & 4024                                                                  & 4024                                                                     & 4024                                                                     \\
      & \begin{tabular}[c]{@{}l@{}}SD\\    \\ 95\% CI\end{tabular} & \begin{tabular}[c]{@{}l@{}}8.963\\    \\ {[}53.70-54.27{]}\end{tabular} & \begin{tabular}[c]{@{}l@{}}8.100\\    \\ {[}14.11-14.60{]}\end{tabular} & \begin{tabular}[c]{@{}l@{}}5.109\\    \\ {[}4.00-4.32{]}\end{tabular} & \begin{tabular}[c]{@{}l@{}}21.120\\    \\ {[}29.81-31.16{]}\end{tabular} & \begin{tabular}[c]{@{}l@{}}22.921\\    \\ {[}70.58-71.95{]} \\
    \end{tabular} \\ \hline \hline
    \label{tab:my_label}
    \end{tabular}
\end{table}

\subsection{Association of breast cancer characteristics by Race and Across Age Groups}
Table \ref{table2} shows the distributions and Chi-square association of race with breast cancer grade, marital status, stage, estrogen/ progesterone, status, and the various categories of survival month, tumor size, RNE, and RNP. Breast cancer grade had a significant association with race (p-value \textless 0.0001), likewise marital status (p-value \textless 0.0001), N stage (p-value = 0.0416), differentiate (p-value \textless 0.0001), estrogen (p-value = 0.0012), status (0.0001), and survival month category (0.0016). T stage (p-value =0.2061), 6th stage (p-value = 0.3560), A stage (p-value = 0.8577), tumor size category (p-value = 0.7481), RNE category (p-value = 0.1462), and RNP category (p-value = 0.8094) were non-significantly associated with race. The distribution across age categories is presented in table \ref{table3}. Grade, marital status, T stage, 6th stage, differentiate, tumor size, estrogen, progesterone, and status (Dead/Alive) were associated across age categories.

\begin{table}[H]
\centering
\caption{Frequency distribution of breast cancer variables by race}
\label{table2}
\begin{tabular}{lllll} \hline  \hline
Characteristics           &              & Race           &                &                    \\
                          & Black n (\%) & Other n   (\%) & White n   (\%) & p-value            \\ \hline
Grade                     &              &                &                &                    \\
1                         & 32(0.80)     & 46 (1.15)      & 465 (11.61)    &                    \\
2                         & 141 (3.52)   & 180 (4.49)     & 2030 (58.70)   & \textless   0.001  \\
3                         & 115 (2.87)   & 94 (2.35)      & 902 (27.74)    &                    \\ \hline
Marital Status            &              &                &                &                    \\
Divorced                  & 40 (0.99)    & 29 (0.72)      & 417 (10.36)    &                    \\
Married                   & 113 (2.81)   & 237 (5.89)     & 2293 (56.98)   &                    \\
Separate                  & 8 (0.20)     & 4 (0.10)       & 33 (0.82)      & \textless   0.001  \\
Single                    & 102 (2.53)   & 33 (0.82)      & 480 (11.93)    &                    \\
Widowed                   & 28 (0.70)    & 17 (0.42)      & 190 (4.72)     &                    \\ \hline
T   Stage                 &              &                &                &                    \\
T1                        & 117 (2.91)   & 113 (2.81)     & 1373 (34.12)   &                    \\
T2                        & 132 (3.28)   & 162 (4.03)     & 1492 (37.08)   & 0.2061             \\
T3                        & 33 (0.82)    & 41 (1.02)      & 459 (11.41)    &                    \\
T4                        & 9 (0.22)     & 4 (0.10)       & 89 (2.21)      &                    \\ \hline
N Stage                   &              &                &                &                    \\
N1                        & 184 (4.57)   & 211 (5.24)     & 2337 (58.08)   &                    \\
N2                        & 63 (1.57)    & 74 (1.84)      & 683 (16.97)    & 0.0416             \\
N3                        & 44 (1.09)    & 35 (0.87)      & 393 (9.77)     &                    \\ \hline
6th Stage                 &              &                &                &                    \\
IIA                       & 91 (2.26)    & 90 (2.24)      & 1124 (27.93)   &                    \\
IIB                       & 81 (2.01)    & 95 (2.36)      & 954 (23.71)    &                    \\
IIIA                      & 69 (1.71)    & 96 (2.39)      & 885 (21.99)    & 0.3560             \\
IIIB                      & 6 (0.15)     & 4 (0.10)       & 57 (1.42)      &                    \\
IIIC                      & 44 (1.09)    & 35 (0.87)      & 393 (9.77)     &                    \\ \hline
Differentiate             &              &                &                &                    \\
Moderately Differentiated & 141 (3.50)   & 180 (4.47)     & 2030 (50.45)   &                    \\
Poorly Differentiated     & 115 (2.86)   & 94 (2.34)      & 902 (22.42)    & \textless 0.0001   \\
Undifferentiated          & 3 (0.07)     & 0 (0.00)       & 16 (0.40)      &                    \\
Well Differentiated       & 32 (0.80)    & 46 (1.14)      & 465 (11.56)    &                    \\ \hline
A Stage                   &              &                &                &                    \\
Regional                  & 283 (7.03)   & 313 (7.78)     & 3336 (82.90)   & 0.8577             \\
Distant                   & 8 (0.20)     & 7 (0.17)       & 77 (1.91)      &                    \\ \hline

\end{tabular}
\end{table}

\begin{table}[t]
\centering
\begin{tabular}{lllll} 

Estrogen                  &              &                &                &                    \\
Negative                  & 33 (0.82)    & 27 (0.67)      & 209 (5.19)     & 0.0012             \\
Positive                  & 258 (6.41)   & 293 (7.28)     & 3204 (79.62)   &                    \\ \hline
Progesterone              &              &                &                &                    \\
Negative                  & 64 (1.59)    & 58 (1.44)      & 576 (14.31)    & 0.0803             \\
Positive                  & 227 (5.64)   & 262 (6.51)     & 2837 (70.50)   &                    \\ \hline
Status                    &              &                &                &                    \\
Alive                     & 218 (5.42)   & 287 (7.13)     & 2903 (72.14)   & \textless   0.0001 \\
Dead                      & 73 (1.81)    & 33 (0.82)      & 510 (12.67)    &                    \\ \hline
Survival Month            &              &                &                &                    \\
$\leq$ 20                      & 20 (0.50)    & 11 (0.27)      & 108 (2.68)     &                    \\
21-49                     & 41 (1.02)    & 30 (0.75)      & 378 (9.39)     &                    \\
50-69                     & 98 (2.44)    & 83 (2.06)      & 1075 (26.71)   & 0.0016             \\
70-89                     & 71 (1.76)    & 109 (2.71)     & 989 (24.58)    &                    \\
$\geq$ 90                      & 61 (1.52)    & 87 (2.16)      & 863 (21.45)    &                    \\ \hline

Tumor Size                &              &                &                &                    \\
$\leq$ 30                      & 192 (4.77)   & 207 (5.14)     & 2269 (56.39)   &                    \\
31-60                     & 75 (1.86)    & 91 (2.26)      & 840 (20.87)    &                    \\
61-90                     & 16 (0.40)    & 17 (0.42)      & 237 (5.89)     & 0.7481             \\
91-120                    & 7 (0.17)     & 4 (0.10)       & 56 (1.39)      &                    \\
$\geq$ 120                     & 1 (0.02)     & 1 (0.02)       & 11 (0.27)      &                    \\ \hline
RNE                       &              &                &                &                    \\
$\leq$ 20                      & 243 (6.04)   & 244 (6.06)     & 2752 (68.39)   &                    \\
21-40                     & 47 (1.17)    & 76 (1.89)      & 654 (16.25)    & 0.1462             \\
$\geq$ 50                      & 1 (0.02)     & 0 (0.00)       & 7 (0.17)       &                    \\ \hline
RNP                       &              &                &                &                    \\
$\leq$ 20                      & 285 (7.08)   & 314 (7.80)     & 3327 (82.68)   &                    \\
21-40                     & 5 (0.12)     & 6 (0.15)       & 79 (1.96)      & 0.8094             \\
$\geq$ 50                      & 1 (0.02)     & 0 (0.00)       & 7 (0.17)       &    \\  \hline  \hline           
\end{tabular} 
\end{table}

\begin{table}[H]
\centering 
\caption{Frequency distribution of breast cancer variables by age group}
\label{table3}
\begin{tabular}{llllll} \hline \hline

Characteristics  &                                                              &              & Age group    &              &                    \\
                 & 30-39                                                        & 40-49        & 50-59        & $\geq$ 60       & p-value            \\ 
                 & n (\%)                                                       & n (\%)       & n (\%)       & n (\%)       &                    \\ \hline 
Grade            &                                                              &              &              &              &                    \\
1                & 12 (0.3)                                                     & 144 (3.6)    & 192 (4.79)   & 195 (4.87)   &                    \\
2                & 120 (3)                                                      & 618 (15.43)  & 837 (20.90)  & 776 (19.38)  & \textless   0.0001 \\
3                & 95 (2.37)                                                    & 357 (8.91)   & 355 (8.86)   & 304 (7.59)   &                    \\ \hline
Race             &                                                              &              &              &              &                    \\
Black            & 24 (0.6)                                                     & 97 (2.41)    & 92 (2.29)    & 78 (1.94)    &                    \\
Other            & 34 (0.84)                                                    & 110 (2.73)   & 93 (2.31)    & 83 (2.06)    & \textless   0.0001 \\
White            & 172 (4.27)                                                   & 917 (22.79)  & 1205 (29.95) & 1119 (27.81) &                    \\ \hline
Marital Status   &                                                              &              &              &              &                    \\
Divorced         & 15 (0.37)                                                    & 121 (3.01)   & 183 (4.55)   & 167 (4.15)   &                    \\
Married          & 146 (3.63)                                                   & 778 (19.33)  & 926 (23.01)  & 793 (19.71)  &                    \\
Separate         & 2 (0.05)                                                     & 19 (0.47)    & 17 (0.42)    & 7 (0.17)     & \textless   0.0001 \\
Single           & 64 (1.59)                                                    & 194 (4.82)   & 209 (5.19)   & 148 (3.68)   &                    \\
Widowed          & 3 (0.07)                                                     & 12 (0.30)    & 55 (1.37)    & 165 (4.10)   &                    \\ \hline

\end{tabular}
\end{table}

\begin{table}[H]
\centering
\begin{tabular}{llllll} 

T   Stage        &                                                              &              &              &              &                    \\
T1               & 75 (1.86)                                                    & 410 (10.19)  & 571 (14.19)  & 547 (13.59)  &                    \\
T2               & 103 (2.56)                                                   & 511 (12.70)  & 607 (15.08)  & 565 (14.04)  & 0.0002             \\
T3               & 45 (1.12)                                                    & 179 (4.45)   & 170 (4.22)   & 139 (3.45)   &                    \\
T4               & 7 (0.17)                                                     & 24 (0.60)    & 42 (1.04)    & 29 (0.72)    &                    \\ \hline
N Stage          &                                                              &              &              &              &                    \\
N1               & 149 (3.70)                                                   & 785 (19.51)  & 916 (22.76)  & 882 (21.92)  &                    \\
N2               & 46 (1.14)                                                    & 223 (5.54)   & 302 (7.50)   & 249 (6.19)   & 0.1908             \\ \hline
6th Stage        &                                                              &              &              &              &                    \\
IIA              & 62 (1.54)                                                    & 345 (8.57)   & 446 (11.08)  & 452 (11.08)  &                    \\
IIB              & 66 (1.64\%)                                                  & 355 (8.82\%) & 364 (9.05)   & 345 (8.57)   &                    \\
IIIA             & 62 (1.54)                                                    & 293 (7.28)   & 378 (9.39)   & 317 (7.88)   & 0.0193             \\
IIIB             & 5 (0.12)                                                     & 15 (0.37)    & 30 (0.75)    & 17 (0.42)    &                    \\
IIIC             & 35 (0.87)                                                    & 116 (2.88)   & 172 (4.27)   & 149 (3.70)   &                    \\ \hline
Differentiate    &                                                              &              &              &              &                    \\
Moderately       & 120 (2.98)                                                   & 618 (15.36)  & 837 (20.80)  & 776 (19.28)  &                    \\
Poorly           & 95 (2.36)                                                    & 357 (8.87)   & 355 (8.82)   & 304 (7.55)   & \textless   0.0001 \\
Undifferentiated & 3 (0.07)                                                     & 5 (0.12)     & 6 (0.15)     & 5 (0.12)     &                    \\
Well             & 12 (0.30)                                                    & 12 (0.30)    & 192 (4.77)   & 195 (4.85)   &                    \\ \hline 
A Stage          &                                                              &              &              &              &                    \\
Regional         & 9 (0.22)                                                     & 32 (0.80)    & 27 (0.67)    & 24 (0.60)    & 0.1111             \\
Distant          & \begin{tabular}[c]{@{}l@{}}221\\    \\ (5.49\%)\end{tabular} & 1092 (27.14) & 1363 (33.87) & 1256 (31.81) &                    \\ \hline
Estrogen         &                                                              &              &              &              &                    \\
Negative         & 31 (0.77)                                                    & 77 (1.91)    & 101 (2.51)   & 60 (1.49)    & \textless   0.0001 \\
Positive         & 199 (4.95)                                                   & 1047 (26.02) & 1289 (32.03) & 1220 (30.32) &                    \\ \hline
Progesterone     &                                                              &              &              &              &                    \\
Negative         & 46 (1.14)                                                    & 156 (3.88)   & 281 (6.98)   & 215 (5.34)   & 0.0003             \\
Positive         & 184 (4.57)                                                   & 968 (24.06)  & 1109 (27.56) & 1065 (26.47) &                    \\ \hline
Status           &                                                              &              &              &              &                    \\
Alive            & 183 (4.55)                                                   & 988 (24.55)  & 1201 (29.85) & 1036 (25.75) & \textless   0.0001 \\
Dead             & 47 (1.17)                                                    & 136 (3.38)   & 189 (4.70)   & 244 (6.06)   &                    \\ \hline
                
                 Survival Month   &                                                              &              &              &              &                    \\
$\leq$ 20             & 11 (0.27)                                                    & 41 (1.02)    & 42 (1.04)    & 45 (1.12)    &                    \\
21-49            & 34 (0.84)                                                    & 113 (2.81)   & 152 (3.78)   & 150 (3.73)   &                    \\
50-69            & 70 (1.74)                                                    & 325 (8.08)   & 448 (11.13)  & 413 (10.26)  & 0.1304             \\
70-89            & 73 (1.81)                                                    & 337 (8.37)   & 400 (9.94)   & 359 (8.92)   &                    \\
$\geq$ 90             & 42 (1.04)                                                    & 308 (7.65)   & 348 (8.65)   & 313 (7.78)   &                    \\ \hline 
Tumor Size       &                                                              &              &              &              &                    \\
$\leq$ 30             & 131 (3.26)                                                   & 709 (17.62)  & 939 (23.33)  & 889 (22.09)  &                    \\
31-60            & 65 (1.62)                                                    & 298 (7.41)   & 349 (8.67)   & 294 (7.31)   &                    \\
61-90            & 28 (0.70)                                                    & 86 (2.14)    & 80 (1.99)    & 76 (1.89)    & 0.0002             \\
91-120           & 6 (0.15)                                                     & 23 (0.57)    & 20 (0.50)    & 18 (0.45)    &                    \\
$\geq$ 120            & 0 (0.00)                                                     & 8 (0.20)     & 2 (0.05)     & 3 (0.07)     &                    \\ \hline 
RNE              &                                                              &              &              &              &                    \\
$\leq$ 20             & 181 (4.50)                                                   & 902 (22.42)  & 1121 (27.86) & 1035 (25.72) &                    \\
21-40            & 49 (1.22)                                                    & 219 (5.44)   & 265 (6.59)   & 244 (6.06)   & 0.8198             \\
$\geq$ 50             & 0 (0.00)                                                     & 3 (0.07)     & 4 (0.10)     & 1 (0.01)     &                    \\ \hline 

\end{tabular}
\end{table}

\begin{table}[H]
\centering
\begin{tabular}{llllll} 
RNP              &                                                              &              &              &              &                    \\
$\leq$ 20             & 225 (5.59)                                                   & 1105 (27.46) & 1349 (33.52) & 1247 (30.99) &                    \\
21-40            & 5 (0.12)                                                     & 16 (0.40)    & 37 (0.92)    & 32 (0.80)    & 0.3059             \\
$\geq$ 50             & 0 (0.00)                                                     & 3 (0.07)     & 4 (0.10)     & 1 (0.02)     &     \\ \hline 
\hline 
\end{tabular}
\end{table}

\subsection{Differences in Age, RNE, RNP, Survival months, and Tumor size by Race across Status}
\begin{figure}[hbt!]
\scalebox{}{}
    \centering
    \includegraphics{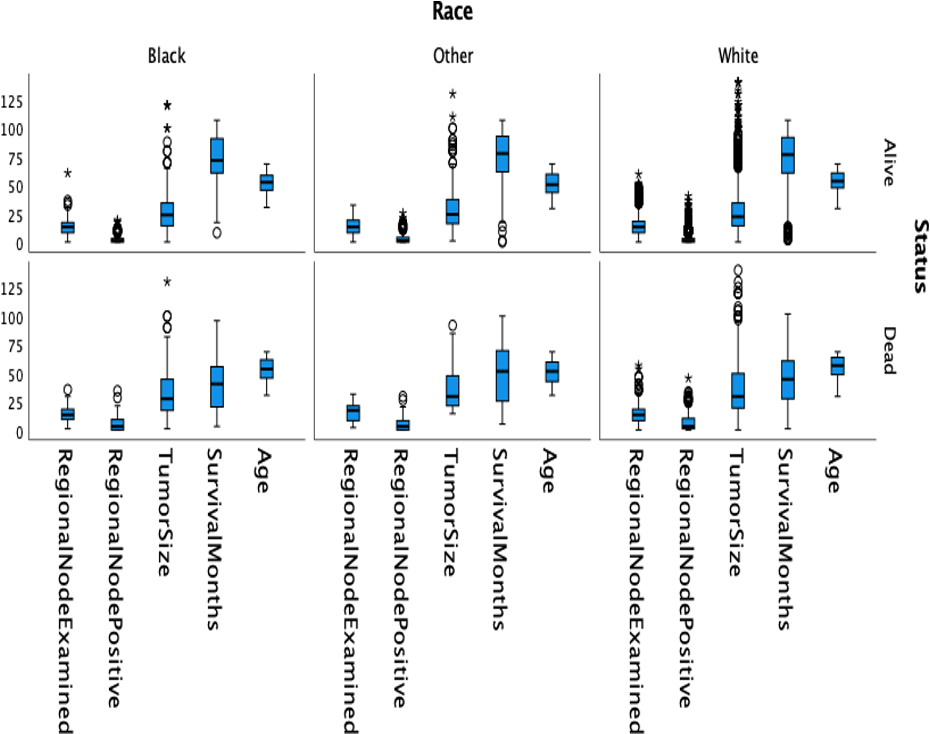}
    \caption{Boxplots of Age, Survival months, RNE, RNP, and Tumor size by Race across status}
    \label{fig:boxplots}
\end{figure}

Figure \ref{fig:boxplots} depicts the side-by-side plot with respect to race across survival status. The mean difference in age between blacks and whites is -1.748 (95\% CI, -2.810 to -0.687), the difference in age between blacks and others (American Indian/AK Native, Asian/Pacific Islander) is 1.150 with 95\% CI from -0.332 to 2.631 and the age difference between white and others is 2.898 (95\% CI, 1.876 to 3.919). For blacks and whites, there was a significant difference in the means of age (p-value = 0.01), and survival months (p-value = 0.001, 95\% CI, -7.866 to -1.945). For blacks and others, the only significant difference was in survival months (p-value \textless 0.001, 95\% CI, -10.415 to -2.793) and for whites and others, the significant difference in mean was recorded for only age (p-value \textless 0.001, 95\% CI, 1.876 to 3.919). There was a significant difference in all but RNE (p-value = 0.027, 95\% CI, 0.087 to 1.477) for status (Dead/Alive). Table \ref{table4} is a summary of the mean differences, p-values, 95\% Confidence Interval (CI), and Cohen’s d effect sizes.

\begin{table}[H]
\centering
\caption{Mean differences between racial combinations and survival status for Age, RNE, RNP, and Survival Months}
\label{table4}
\begin{tabular}{llllll} \hline \hline
Characteristics &                 & Mean Difference & p-value          & 95\% CI               & Cohen’s d \\ \hline 
Age             & Black vs Whites & -1.748          & 0.001            & {[}2.810, -0.687{]}   & -0. 197   \\
                & Black vs Other  & 1.150           & 0.128            & {[}-1.593, 0.793{]}   & 0.123     \\
                & White vs Other  & 2.898           & \textless{}0.001 & {[}1.876, 3.919{]}    & 0.325     \\
Status          & Dead vs   Alive & 1.392           & \textless 0.001  & {[}0.624, 2.160{]}{]} & 0.156     \\ \hline
RNE             & Black vs Whites & -0.015          & 0.976            & {[}-0.991, 0.961{]}   & -0.002    \\
                & Black vs Other  & -0.400          & 0.511            & {[}-1.593, 0.793{]}   & -0.053    \\
                & White vs Other  & -0.385          & 0.419            & {[}-1.319, 0.549{]}   & -0.047    \\ 
Status          & Dead vs   Alive & 0.782           & 0.027            & {[}0.087, 1.477{]}    & 0.097     \\  \hline
RNP             & Black vs Whites & 0.292           & 0.352            & {[}-3.22, 0.906{]}    & 0.057     \\
                & Black vs Other  & 0.216           & 0.592            & {[}-0.577, 1.010{]}   & 0.043     \\
                & White vs Other  & -0.075          & 0.801            & {[}-0.661, 0.511{]}   & -0.015    \\
Status          & Dead vs   Alive & 3.641           & \textless 0.001  & {[}3.217, 4.065{]}    & 0.737     \\ \hline
Tumor size      & Black vs Whites & 0.102           & 0.937            & {[}-2.440, 2.643{]}   & 0.005     \\
                & Black vs Other  & -0.475          & 0.780            & {[}-3.816, 2.867{]}   & -0.023    \\
                & White vs Other  & -0.576          & 0.639            & {[}-2.988, 1.835{]}   & -0.027    \\
Status          & Dead vs   Alive & 7.871           & \textless{}0.001 & {[}6.074, 9.667{]}    & 0.376   \\  \hline \hline
\end{tabular} 
\end{table}

\subsection{Effects of Grade, Race, T stage, N stage, Estrogen, Progesterone, Age, RNE, and RNP on Status}
The effects of grade, race, T stage, N stage, estrogen, progesterone, age, RNE, and RNP were p- value = \textless{0.0001}  for grade 1 and 0.0002 for grade 2, p-value= 0.0004 for black and 0.0332 for others (American Indian/AK Native, Asian/Pacific Islander), p-value = \textless{0.0001} for T1, 0.002 for T2 and 0.0191 for T3, p-value = 0.0125 for N1 and 0.3817 for N2, p-value \textless {0.001} for estrogen and progesterone (negative), p-value \textless {0.001} for age, RNE, and RNP. The odds ratio for grades 1 and 3 is 68\% high (2.488, [CI, 1.709 to 3.622]) than for grades 2 and 3 (1.477, [CI, 1.203 to 1.813]). The racial odds ratio for black and white was lower (0.566, [CI, 0.414 to 0.773]) compared to other (American Indian/AK Native, Asian/Pacific Islander) and white (1.537, [CI, 1.035 to 2.283]). For the T stage, the odds ratios were T1 vs T4 (3.136, [1.943 - 5.060]), T2 vs T4 (2.066, [CI, 1.303 - 3.276]), and T3 vs T4 (1.806, [CI, 1.101 - 2.960]) and the odds for N stage (N1 vs N2 and N2 vs N3) are 1.798 [CI, 1.135 - 2.849], and 1.189 [CI, 0.807 - 1.751] respectively. The odds for estrogen (negative vs positive) and progesterone (negative vs positive) were 0.484 [CI, 0.341 - 0.687] and 0.552 [CI, 0.430 - 0.710] respectively. Table \ref{table5} shows the logistic regression estimates for the 4024 sampled women.

\begin{table}[H]
\centering
\caption{Logistic regression estimates}
\label{table5}
\begin{tabular}{llllll} \hline \hline
Characteristics & Estimates & SE      & Effect         & OR (95\% Wald CI)       & p-value            \\
Grade           &           &         &                &                         &                    \\
1               & 0.9116    & 0.1916  & Grade 1 vs 3   & 2.488 (1.709 - 3.622)   & \textless   0.0001 \\
2               & 0.3898    & 0.1047  & Grade 2 vs 3   & 1.477 (1.203 - 1.813)   & 0.0002             \\
3               & Ref       &         &                &                         &                    \\ \hline
Race            &           &         &                &                         &                    \\
Black           & -0.5693   & 0.1593  & Black vs White & 0.566 (0.414 - 0.773)   & 0.0004             \\
Other           & 0.4300    & 0.2019  & Other vs White & 1.537 (1.035 - 2.283)   & 0.0332             \\
White           & Ref       &         &                &                         &                    \\ \hline
T   Stage       &           &         &                &                         &                    \\
T1              & 1.1429    & 0.2441  & T1 vs T4       & 3.136 (1.943 - 5.060)   & \textless   0.0001 \\
T2              & 0.7255    & 0.2352  & T2 vs T4       & 2.066 (1.303 - 3.276)   & 0.0002             \\
T3              & 0.5909    & 0.2522  & T3 vs T4       & 1.806 (1.101 - 2.960)   & 0.0191             \\
T4              & Ref       &         &                &                         &                    \\ \hline

\end{tabular}
\end{table}

\begin{table}[H]
\centering
\begin{tabular}{llllll} 
N Stage         &           &         &                &                         &                    \\
N1              & 0.5867)   & 0.2349  & N1 vs N3       & 1.798   (1.135 - 2.849) & 0.0125             \\
N2              & 0.1728    & 0.1976  & N2 vs N3       & 1.189 (0.807 - 1.751)   & 0.3817             \\
N3              & Ref       &         &                &                         &                    \\ \hline
Estrogen        &           &         &                &                         &                    \\
Negative        & -0.7259   & 0.1786  & Neg vs Pos     & 0.484 (0.341 - 0.687)   & \textless   0.001  \\
Positive        & Ref       &         &                &                         &                    \\ \hline
Progesterone    &           &         &                &                         &                    \\
Negative        & -0.5935   & 0.1279  & Neg vs Pos     & 0.552 (0.430 - 0.710)   & \textless   0.001  \\
Positive        & Ref       &         &                &                         &                    \\ \hline
Age             & -0.0240   & 0.00547 &                & 0.976 (0.966 - 0.987)   & \textless   0.001  \\ \hline
RNE             & 0.0370    & 0.00726 &                & 1.038 (1.023 - 1.053)   & \textless   0.001  \\
RNP             & -0.0814   & 0.0153  &                & 0.922 (0.895 - 0.950)   & \textless 0.001   \\ \hline \hline
\end{tabular}
\end{table}

\section{Discussion}

The findings from this study indicate cancer grade, marital status, N stage, differentiation, estrogen, and status (dead/ Alive) have a statistical association with breast cancer with the distribution of race groups being predominantly White with Blacks having the least representation similar to the other results having an overwhelming majority of white patients in the sample \cite{stapleton2018raceethnicity}. This may be due to racial disparities in access to healthcare facilities, inadequate health insurance, and behavioral factors which have been captured to be some of the contributing factors by \cite{arciero2017africanamerican, williams2017disparities}. These significant associations give an insight into the characteristics of breast cancer that can be used to predict survival months, grade, tumor size, and other related breast cancer outcomes. The associations between T stage, 6th stage, A stage, and progesterone were statistically non-significant suggesting that racial distribution across these factors does not differ. However, further research would have to be conducted to eliminate potential confounders to make a more accurate conclusion.
 
In table \ref{table1} we present our findings on the average differences in age, RNE, RNP, tumor size, and survival months. The mean age for Blacks was significantly lower than for Whites whereas Whites had a significantly higher average age at diagnosis than for Others. Blacks on average had significantly lower survival months compared to Whites and Others while Others had significantly higher survival months than Whites. These findings correlate with the findings of \cite{hendrick2021agedistributions}. Another important finding is that there was no significant difference in racial groups and status (Dead/ Alive) for RNE, RNP, and tumor size suggesting that the number of RNE, RNP, and tumor size are about the same for surviving or dead patients and across racial groups at diagnostics.
The logistic regression estimates presented in table \ref{table5} provide a useful overview of the factors that are significantly associated with breast cancer. For age, the log odds of dying from a breast cancer-related disease increase by 2.1\% whereas for negative progesterone and estrogen status, the odds decrease by 96.9\% and 70.1\% respectively compared to a positive status. The log odds for black mortality increase relative to whites compared to the odds of others (American Indian/AK Native, and Asian/Pacific Islander) who has a decreasing log odds of mortality compared to the white race.
\section{Limitation}

A limitation of this study was the unavailability of further information such as income, and demographic indicators that could explain some of the disparities from the findings in the study.

\section{Conclusion}
 
In conclusion, the presented results provide important information on the distribution and association of breast cancer characteristics among different races and age groups. These findings highlight the need for further research to better understand the underlying causes of breast cancer disparities and to develop targeted interventions to improve breast cancer outcomes.


\section*{Acknowledgments}

I would like to express my sincere gratitude to Professor Rana Jaber for her invaluable guidance, support, and mentorship throughout the course of this project. Her expertise, encouragement, and constructive feedback have been instrumental in shaping this work and my academic journey. I am deeply thankful for her dedication and inspiration.


\bibliographystyle{myjmva}
\bibliography{trial}
\end{document}